\documentclass[pre,twocolumn,letterpaper,amsfonts,superscriptaddress]{revtex4}
\usepackage{graphicx}

\usepackage{amsthm}
\usepackage{amsmath}
\usepackage{amsfonts}

\usepackage{amssymb}

\usepackage{bm}

\begin{document}

\bibliographystyle{apsrev}
% to cite, use command: \cite{...reference tag...}
% in Bibliography section, enter a citation using the command:
% \bibitem{...reference tag...} then type away.

\title{Rapidly detecting disorder in rhythmic biological signals:
A spectral entropy measure to identify cardiac arrhythmias}

\author{Phillip P.A. Staniczenko}
%\email{phillip.staniczenko@physics.ox.ac.uk}
\affiliation{Physics Department, Clarendon Laboratory, Oxford University, Oxford OX1 3PU, UK}
\affiliation{CABDyN Complexity Centre, Oxford University, Oxford OX1 1HP, UK}
\author{Chiu Fan Lee}
\affiliation{Physics Department, Clarendon Laboratory, Oxford University, Oxford OX1 3PU, UK}
\affiliation{CABDyN Complexity Centre, Oxford University, Oxford OX1 1HP, UK}
\author{Nick S. Jones}
\affiliation{Physics Department, Clarendon Laboratory, Oxford University, Oxford OX1 3PU, UK}
\affiliation{CABDyN Complexity Centre, Oxford University, Oxford OX1 1HP, UK}
\affiliation{Oxford Centre for Integrative Systems Biology, Oxford OX1 3QU, UK}

\date{\today}

\begin{abstract}

We consider the use of a running measure of power spectrum disorder
to distinguish between the normal sinus rhythm of the heart and two forms of cardiac arrhythmia:
atrial fibrillation and atrial flutter.
This spectral entropy measure is motivated by characteristic differences in the power spectra of beat timings
during the three rhythms.
We plot patient data derived from ten-beat windows on a ``disorder map" and identify
rhythm-defining ranges in the level and variance of spectral entropy values.
Employing the spectral entropy within an automatic arrhythmia
detection algorithm enables the classification of periods of atrial fibrillation
from the time series of patients' beats.
When the algorithm is set to identify abnormal rhythms within 6 s
it agrees with 85.7\% of the annotations of professional rhythm assessors;
for a response time of 30 s this becomes 89.5\%, and with 60 s it is 90.3\%.
The algorithm provides a rapid way to detect atrial fibrillation,
demonstrating usable response times as low as 6 s.
Measures of disorder in the frequency domain have practical significance in
a range of biological signals: the techniques described in this paper have
potential application for the rapid identification of disorder in other rhythmic signals.

\end{abstract}

\pacs{Put PACS numbers here} % put showpacs in documentclass to display

\maketitle

% Introduction
\section{Introduction}
\label{Introduction}
% Set scene
Cardiovascular diseases are a group of disorders of the heart
and blood vessels and are the largest cause of death globally \cite{WHO}.
An arrhythmia is a disturbance in the normal rhythm of
the heart and can be caused by a range of cardiovascular diseases.
% Atrial fibrillation in particular
In particular, atrial fibrillation is a common arrhythmia
affecting 0.4\% of the population and 5\%-10\%
of those over 60 years old \cite{Kannel}; it can lead
to a very high (up to 15-fold) risk of stroke \cite{Bennett}.
Heart arrhythmias are thus a clinically significant domain
in which to apply tools investigating the dynamics of
complex biological systems \cite{Wessel}.
% Placing us
Since the pioneering work of Akselrod \emph{et al.} on spectral aspects of
heart rate variability \cite{Akselrod}, such approaches
have tended to focus on frequencies lower than the breathing rate.
By contrast, we develop a spectral entropy measure to investigate
heart rhythms at higher frequencies, similar to the heart
rate itself, that can be meaningfully applied to short segments of data.

% Other methods
Conventional physiological measures of disorder,
such as approximate entropy (ApEn) and sample entropy (SampEn),
typically consider long time series as a whole
and require many data points to give useful results \cite{Grassberger}.
With current implant technology
and the increasing availability of portable electrocardiogram (ECG) devices \cite{PortECG},
a rapid approach to fibrillation detection is both possible and sought after.
Though numerous papers propose rapid methods for detecting
atrial fibrillation using the ECG (\cite{Xu} and refs therein),
less work has been done using only the time series of beats or intervals
between beats (\emph{RR} intervals).
In one study, Tateno and Glass use a statistical method comparing standard
density histograms of \emph{$\Delta$RR} intervals \cite{TatenoGlass}.
The method requires around 100 intervals to detect a change in behavior
and thus may not be a tool suitable for rapid response.

% Application to other biological systems
Measures of disorder in the frequency domain have practical significance in
a range of biological signals.
The irregularity of electroencephalography (EEG) measurements
in brain activity, quantified using the entropy of the power spectrum,
has been suggested to investigate localized
desynchronization during some mental and motor tasks \cite{Inouye}.
Thus, the techniques described here have potential application
for the rapid identification of disorder in
other rhythmic signals.

% Outline
In this paper we present a technique for quickly quantifying
disorder in high frequency event series:
the spectral entropy is a measure of disorder applied to the
power spectrum of periods of time series data.
By plotting patient data on a disorder map, we observe distinct
thresholds in the level and variance of spectral entropy values
that distinguish normal sinus rhythm from two arrhythmias:
atrial fibrillation and atrial flutter.
We use these thresholds in an algorithm designed to automatically
detect the presence of atrial fibrillation in patient data.
When the algorithm is set to identify abnormal rhythms within 6 s
it agrees with 85.7\% of the annotations of professional rhythm assessors;
for a response time of 30 s this becomes 89.5\%, and with 60 s it is 90.3\%.
The algorithm provides a rapid way to detect fibrillation,
demonstrating usable response times as low as 6 s
and may complement other detection techniques.

% Outline of sections to come
The structure of the paper is as follows.
Section \ref{Analysis} introduces the data analysis and methods employed in the
arrhythmia detection algorithm,
including a description of the spectral entropy and disorder map
in the context of cardiac data.
The algorithm itself is presented in Sec. \ref{Algorithm}, along with results for a
range of detection response times.
In Sec. \ref{Discussion}, we discuss the results of the algorithm and sources of error, and compare our method to other fibrillation detection techniques.
An outline of further work is presented in Sec. \ref{FurtherWork},
with a summary of our conclusions closing the paper in Sec.
\ref{Conclusion}.

% One line leeway.

\section{Data Analysis}
\label{Analysis}

After explaining how we symbolize cardiac data in Sec. \ref{SymbolizingCardiacData},
the spectral entropy measure is introduced (Sec. \ref{SpectralEntropy}) and
appropriate parameters for cardiac data are selected (Sec. \ref{ParameterSelection}).
We then show how the various rhythms of the
heart can be identified by their position on a disorder map defined
by the level and variance of spectral entropy values (Sec. \ref{CardiacRhythmSpace}).

% Raw data
Data were obtained from the MIT-BIH atrial fibrillation database (afdb),
which is part of the \emph{physionet} resource \cite{physionet}.
This database contains 299 episodes of atrial fibrillation and 13
episodes of atrial flutter across 25 subjects (henceforth referred to as ``patients"),
where each patient's Holter tape
is sampled at 250 Hz for 10 h. The onset and end of atrial fibrillation
and flutter were annotated by trained observers as part of the database.
The timing of each QRS complex (denoting contraction of the ventricles
and hence a single, ``normal", beat of the heart)
had previously been determined by an automatic detector \cite{Laguna}.

\subsection{Symbolizing cardiac data}
\label{SymbolizingCardiacData}

\begin{figure}[b]
\centering
\includegraphics[width=0.45\textwidth]{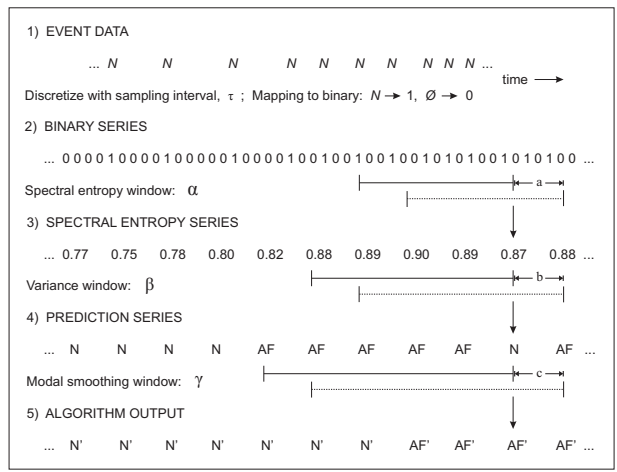}
\caption[Arrhythmia Detection Algorithm Schematic]
{
Schematic of cardiac data analysis and the automatic arrhythmia detection algorithm.
A full description of the data analysis (stages 1-3) is given in the
Data Analysis section, Sec. \ref{Analysis}, of the text; the remaining steps (stages 4-5) are
described in the Algorithm section, Sec. \ref{Algorithm}.
MIT-BIH atrial fibrillation database event data (stage 1) are
discretized at sampling interval $\tau$, then mapped to give
a binary series representing the dynamics of regular beats \emph{N} (stage 2).
A sequence of spectral entropy windows, of length $\alpha$, is applied with
overlap parameter \emph{a} to obtain a series of spectral entropy values (stage 3).
Variance windows, of length $\beta$ with overlap parameter \emph{b},
are applied to obtain a series of variance values.
Threshold conditions in the level and variance of spectral entropy values
allows for the classification of periods of atrial fibrillation (AF) and
other rhythms (N), typically normal sinus rhythm (stage 4).
Finally, the most frequent prediction in each modal smoothing window,
of length $\gamma$ with overlap parameter \emph{c},
is identified \{AF$^{\prime}$, N$^{\prime}$\}
to obtain the final algorithm output (stage 5).
For definitions and typical values for algorithm parameters,
see Table \ref{taba}.
}
\label{Figure:Schematic}
\end{figure}

% Symbolic string
We convert event data into a binary
string, a form appropriate for use in the spectral entropy measure.
The beat data is an event series: a sequence of pairs denoting the
time of a beat event and its type. We categorize normal beats as
\emph{N} and discretize time into short intervals of
length $\tau$ (for future reference, symbols are collected with
summarizing descriptions in Table \ref{taba}).
Each interval is categorized as \emph{\O} or \emph{N} depending
on whether it contains no recorded event or a normal beat, respectively.
This yields a symbolic string of the form
...\emph{\O}\emph{\O}\emph{\O}\emph{N}\emph{\O}\emph{\O}\emph{N}\emph{\O}\emph{N}\emph{\O}\emph{\O}\emph{\O}\emph{N}....
This symbolic string can be mapped to a binary sequence (\emph{N} $\to$ 1, \emph{\O} $\to$ 0).
This procedure is shown schematically in Fig. \ref{Figure:Schematic}.
Naturally, this categorization can be extended to more than two states and
applied to other systems.
For example, ectopic beats (premature ventricular contractions) could
be represented by \emph{V} to yield a symbolic string drawn from the set
\{\emph{\O},\emph{N},\emph{V}\}. An additional map could then be used to
extract a binary string representing the dynamics of ectopic beats.

\subsection{Spectral entropy}
\label{SpectralEntropy}

% Physiological motivation
We now present a physiological motivation for using a measure of disorder
in the context of cardiac dynamics, followed by a description of the spectral entropy measure.
Following Ref. \cite{Bennett}, atrial fibrillation is characterized
by the physiological process of \emph{concealed conduction} in which
the initial regular electrical impulses from the atria (upper chamber of the heart)
are conducted intermittently by the atrioventricular node to the ventricles
(lower chamber of the heart).
This process is responsible for the irregular ventricular rhythm that is observed.
Atrial flutter has similar causes to atrial fibrillation but is less common;
incidences of flutter can degenerate into periods of fibrillation.
Commonly, alternate electrical waves are conducted to the ventricles,
maintaining the initial regular impulses originating from the atria.
This results in a rhythm with pronounced regularity.
Normal sinus rhythm can be characterized by a slightly less regular beating
pattern occurring at a slower rate compared to atrial flutter.
% Reference ECG traces
Example electrocardiograms for the three rhythms are shown in the boxed-out areas of Fig. \ref{Figure:Phase_Space}, below.

Given these physiological phenomena, the spectral entropy can be used as a
natural measure of disorder, enabling one to distinguish between these three
rhythms of the heart.
Presented with a possibly very short period of heart activity one can create a length-$L$,
duration-$L\tau$, binary string.
One then obtains the corresponding power spectrum of this period of heart activity using
the discrete Fourier transform \cite{DFT}.
Given a (discrete) power spectrum with the \emph{i}th
frequency having power \emph{C$_{i}$}, one can define the
``probability" of having power at this frequency as
\begin{eqnarray}
p_{i}=\frac{C_i}{\sum_{i}{C_i}}.
\label{power_eqn}
\end{eqnarray}
When employing the discrete Fourier transform, the summation runs from $i=1$ to $i=\frac{L}{2}$.
One can then find the entropy of this probability distribution
[with \emph{i} having the same summation limits as in Eq. (\ref{power_eqn})]:
\begin{eqnarray}
H=\sum_{i}-p_{i}\log_{2}p_{i}.
\label{H_eqn}
\end{eqnarray}
Breaking the time series into many such blocks of duration $L\tau$,
each with its own spectral entropy, thus returns a time series of spectral entropies.
Note that this measure is not cardiac specific and can be applied to any event series.
% How spectral entropy works
For example, a sine wave having period an integer fraction of $L\tau$ will be
represented in Fourier space by a delta function (for $L\tau\to\infty$) centered at its fundamental
frequency; this gives the minimal value for the spectral entropy of zero.
Other similar frequency profiles, with power located at very specific frequencies,
will lead to correspondingly low values for the spectral entropy.
By contrast, a true white noise signal will have power at all frequencies,
leading to a flat power spectrum. This case results in the maximum value
for the spectral entropy:
\begin{eqnarray}
H_{max} &=&  \log_{2} \left(\frac{L}{2} \right).
\label{Hmax_eqn}
\end{eqnarray}
As will be discussed in the following section, $H$ can be
normalized by $H_{max}$ to give spectral entropy values in
the range [0,1].

% Chiu Fan's stuff
Note that analytical tools relying on various interbeat intervals
have been devised in the past (e.g. \cite{TatenoGlass,SchulteFrohlinde,Lerma}).
Here, we demonstrate how our measure relates to those studies.
Any series of events can be represented by
\begin{eqnarray}
f(t) = \sum_k \delta(t-t_k),
\end{eqnarray}
where $t_k$ is the time when an event (beat) occurs.
The corresponding  power spectrum is, then,
\begin{eqnarray}
P(\omega) \propto
\sum_{k,k'} \cos \big( \omega |t_k-t_{k'}|\big).
\label{cos_eqn}
\end{eqnarray}
The spectral entropy is, by definition,
\begin{eqnarray}
H_{cont.} = \int {\rm d} \omega \, p(\omega) \log p(\omega),
\label{content}
\end{eqnarray}
where $p(\omega) = P(\omega) /\int {\rm d} \omega' P(\omega')$.
We therefore see that  Eq. (\ref{content}) depends on all of the intervals
between \emph{any} two events (c.f. Eq. (\ref{cos_eqn})).
This is in contrast to studies on the distribution of beat-next-beat
intervals in \cite{SchulteFrohlinde}.
We believe that this generalization enriches the structure captured in the short-time segments
and thus allows for the shortening of the detection response time in our
arrhythmia detection algorithm.
We finally note that since the spectral entropy depends only on the \emph{shape}
of the power spectrum, it is relatively insensitive to small, global, shifts
in the spectrum of the signal.

% FIGURE: SCHEMATIC

% TABLE: WINDOW SUMMARY
\begin{table*}
\centerline{
\begin{tabular}{|c|c|c|c|c|c|}
\hline
Window&Symbol&Definition&Typical value&Overlap&Typical value\\
\hline
Spectral entropy&$\alpha$&$L\tau$&6 s&\emph{a}&1.5 s\\
Variance&$\beta$&$Ma=M L\tau/4$&6 s, 30 s, 60 s&\emph{b}&1.5 s\\
Modal smoothing&$\gamma$&$2\beta + \emph{b}=(2M + 1) L\tau/4$&12 s, 60 s, 120 s&\emph{c}&1.5 s\\
\hline
\end{tabular}}
\caption{Summary of arrhythmia detection algorithm window and overlap symbols.
A full description of the spectral entropy and variance windows is given in the
Data Analysis section, Sec. \ref{Analysis}, of the text; the modal smoothing window
is described in the Algorithm section, Sec. \ref{Algorithm}.
Cardiac data in the MIT-BIH atrial fibrillation database is sampled
at intervals of $\tau$=30 ms.
The number of intervals contained in the spectral entropy window, $L$,
is chosen for each patient such that the spectral entropy window is expected to contain ten beats.
In the variance window, $M$ represents the number of spectral entropy values
used in finding the variance; for response times
6 s, 30 s, 60 s, we consider $M$ equal to 4, 20, 40, respectively.
Specifying $\tau$, $L$ and $M$ fixes the remaining parameters.
We define overlap parameter \emph{a}=$\alpha/4$.
For simplicity, we set \emph{c}=\emph{b}=\emph{a}.
}
\label{taba}
\end{table*}

\subsection{Parameter selection}
\label{ParameterSelection}

We now introduce parameters for the spectral entropy measure and
select values appropriate for cardiac data.
% Tau
The sampling interval acts like a low pass-filter on the data
since all details at frequencies above $1/(2\tau)$ Hz, the upper frequency limit,
are discarded \cite{deBoer}.
The sampling interval must be sufficiently small such that no useful
high-frequency components are lost.
We choose a sampling interval $\tau$=30 ms,
since processes like the heart beat interval, breathing and blood pressure
fluctuations occur at much lower frequencies.
The upper frequency limit in the power spectrum
is consistent with the inclusion of all dominant and subsidiary frequency
peaks present during atrial fibrillation \cite{Goldberger}.

% Collapse window size
We call the duration over which the power spectrum is found, and hence a single
spectral entropy value is obtained, the spectral entropy window: $\alpha=L\tau$
($L$ is the number of sampling intervals required).
With our value for $\tau$, the shortest spectral entropy window giving sufficient resolution
in the frequency domain for cardiac data is found for $L$ around 200, $\alpha\sim6$ s.
This value for $\alpha$ is equivalent to approximately ten beats on average
over the entire afdb.
It is consistent with previous work on animal hearts looking
at the minimum window length required to determine values for the
dominant frequencies present during atrial fibrillation \cite{Haines}.
% HR normalization
To take into account the heterogeneity of patients' resting heart rates (HRs),
we fix $\tau$ and use an $L$ value for each patient such that there are on average
10 beats in each spectral entropy window.
Thus, $\alpha=L(\overline{HR})\tau=\alpha(\overline{HR})$.
All subsequent parameters that are determined by $L$ will similarly
be a function of the average heart rate;
we will henceforth drop the $\overline{HR}$ notation for clarity,
with the dependence on average heart rate understood implicitly.
Patients with higher average heart rate require smaller $L$,
and therefore have a shorter spectral entropy window.
% Renormalization of spectral entropy
By invoking individual values for $L$, the maximum spectral entropy
for each patient is constrained to a particular value: $H_{max}$ [c.f. Eq. (\ref{Hmax_eqn})].
To make spectral entropy values comparable, we normalize the
basic spectral entropy values for each patient [Eq. (\ref{H_eqn})]
by their theoretically maximal spectral entropy value.
The spectral entropy can thus take values in the range [0,1].
% Rectangular window
In choosing $L$ near its minimally usable value, we necessarily have
a small number of beats compared to the window length $\alpha$.
In such cases, a window shape having a low value for the equivalent noise bandwidth (ENBW)
is preferable \cite{Harris}.
The ENBW is a measure of the noise associated with a particular window shape:
it is defined as the width of a fictitious rectangular filter such that power
in that rectangular band is equal to the actual power of the signal.
The condition for low ENBW is satisfied by the \emph{rectangular window}.
% slidejump parameter
To maximize the available data, a sequence of overlapping rectangular windows separated
by a time \emph{a} is used.
This results in a series of spectral entropy values also separated by \emph{a}.
We follow the convention of using adjacent window overlap of 75\% \cite{Harris},
leading to a window separation time: $a=L\tau/4$.
This gives a typical value for \emph{a} of 1.5 s.
A summary of window and overlap parameters is presented in Table \ref{taba}.

% FIGURE: PATIENT 08378
\begin{figure}[b]
\centering
\includegraphics[width=0.5\textwidth]{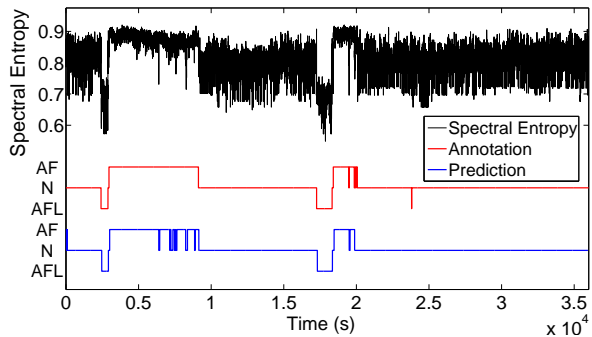}
\caption[Spectral entropy for Patient 08378.]
{
(Color online) Spectral entropy time series (top), professional rhythm annotation (middle),
and arrhythmia detection algorithm prediction (bottom) for patient 08378 from the MIT-BIH
atrial fibrillation database.
Event data is sampled at 30-ms intervals approximately 200 times
such that there are on average ten beats
per spectral entropy window. Each window, of length 6 s for a typical patient,
contributes one value of the spectral entropy;
windows have a typical overlap of 1.5 s.
For the rhythm annotation and algorithm prediction:
AF denotes atrial fibrillation, AFL denotes atrial flutter, and
N represents all other rhythms.
The algorithm prediction (primed symbols omitted for clarity) demonstrates good agreement with professional annotations;
shown for a response time of 30 s, thresholds:
$\Gamma_{fib}$=0.84, $\Gamma_{fl}$=0.70 and $\Phi_{fib}$=0.018.
}
\label{Figure:Patient_08378}
\end{figure}

% FIGURE1: Single spectral entropy trace, comments
Figure \ref{Figure:Patient_08378}
illustrates the spectral entropy measure applied to patient 08378 from the afdb.
We identify three distinct levels in the spectral entropy value
corresponding to the three rhythms of the heart assessed in the annotations.
Beating with a relatively regular pattern, which can be associated with
normal sinus rhythm, sets a baseline for the spectral entropy.
The irregularity associated with fibrillation causes an increase in the value,
with the pronounced regularity of flutter identifiable as a decrease
in the spectral entropy.
% Comment on HR-only measure
We note that power spectrum profiles in frequency space should remain qualitatively similar
for a given rhythm type, regardless of the underlying heart rate.
For example, periodic signals can be characterized by peaks at
constituent frequencies, independent of the beat production rate;
similarly, highly disordered signals can be consistently identifiable by
their flat power spectra.
This confers a significant advantage over methods relying solely on the heart rate.
We find considering only the instantaneous heart rate and its derivatives to
be insufficient in consistently distinguishing between sinus rhythm,
fibrillation and flutter; this point is addressed further in the
Discussion section (Sec. \ref{DisagreementswithAnnotations}).

\subsection{Cardiac disorder map}
\label{CardiacRhythmSpace}

% FIGURE: CARDIAC DISORDER MAP
\begin{figure*}
\centering
\includegraphics[width=0.982\textwidth]{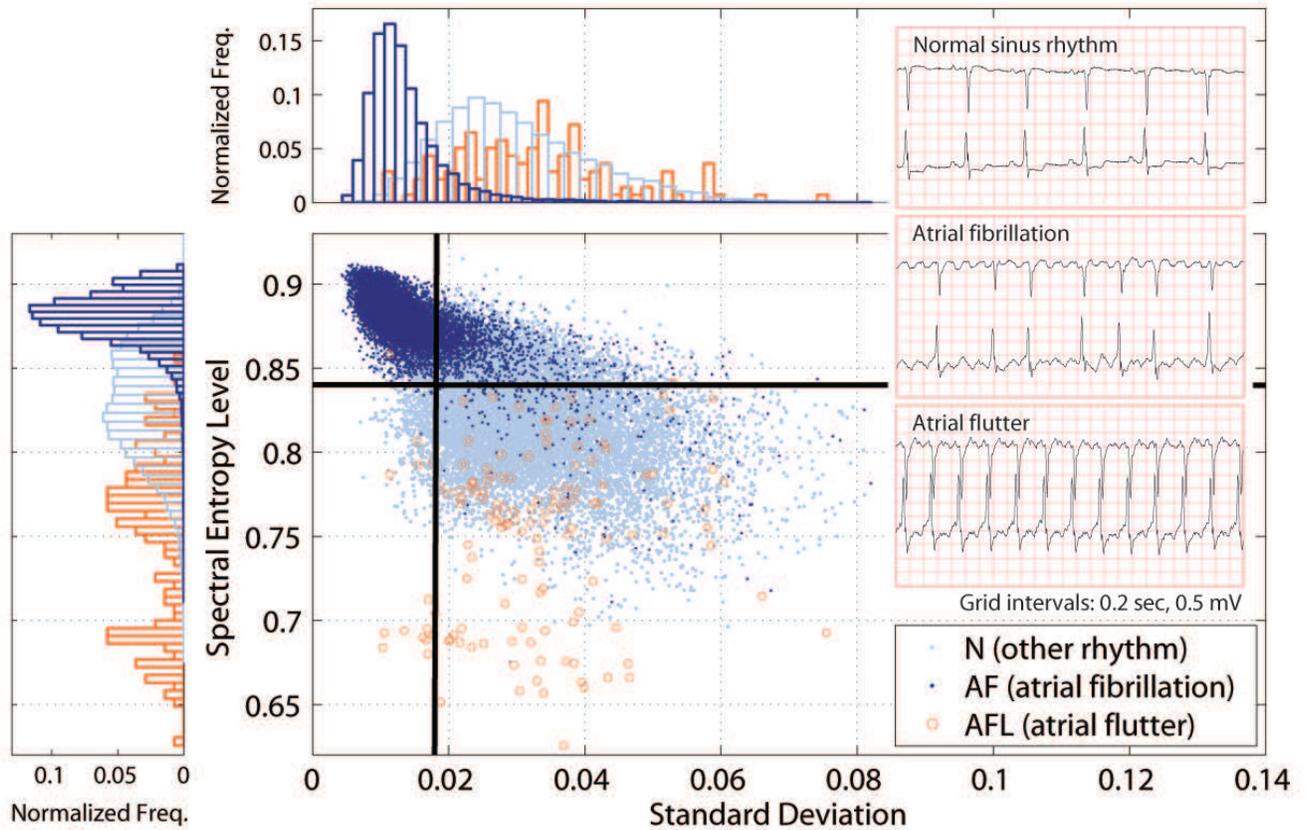}
\caption[Cardiac Disorder Map.]
{
(Color online) Cardiac disorder map for all 25 patients in the MIT-BIH
atrial fibrillation database (afdb).
Boxed-out area: example electrocardiograms for normal sinus rhythm, atrial fibrillation and
atrial flutter, taken from patient 04936.
Spectral entropy values are obtained from windows of event data
expected to contain ten beats; data is sampled at 30-ms intervals.
For a typical patient, each spectral entropy window is around 6 s in length and
has an overlap with adjacent windows of 1.5 s.
For each point on the disorder map, the standard deviation and average
spectral entropy level is calculated from $M$ adjacent spectral entropy values:
we call this the variance window, $\beta$.
Here, we have $M$ equal to 20 and so $\beta$ has a typical
length of 30 s; $\beta$ represents the response time
of the algorithm.
Normalized frequency histograms are disorder map projections onto the relevant axes.
Rhythm assessments, \{N, AF, AFL\}, are provided in the afdb.
Atrial fibrillation is situated in the upper left of the disorder map,
consistent with having a high value for the spectral entropy and a low value
for the variance. Atrial flutter has a lower average value for
the spectral entropy, as expected.
Fibrillation thresholds for the arrhythmia detection algorithm are set at
$\Gamma_{fib}$=0.84 for the spectral entropy level and
$\Phi_{fib}$=0.018 for the standard deviation, as indicated on the disorder map.
}
\label{Figure:Phase_Space}
\end{figure*}

% Why difference in variance of SE values
Having identified differences in the level of the spectral entropy measure
corresponding to different rhythms of the heart, we suggest that there
should be a similar distinction in the \emph{variance} of a series of
spectral entropy values.
We propose that the fibrillating state
may represent an upper limit to the spectral entropy measure;
once this state is reached, variations in the measure's value are
unlikely until a new rhythm is established.
By contrast, the beating pattern of normal sinus rhythm is not as disordered
as possible and can therefore show variation in the spectral entropy values taken.
Inspecting the data, one frequently observes transitions between periods
of very regular and more irregular (though still clearly sinus) beating.
Thus, normal sinus rhythm will naturally explore more of the spectral entropy
value range than atrial fibrillation, which is consistently irregular in character
(including some dominant frequencies \cite{Goldberger}).
Furthermore, in defining the spectral entropy window to be constant for a
given patient, some dependence on the heart rate is retained,
despite accounting for each patient's average heart rate.
This dependence can cause additional harmonics to appear in the power spectrum,
increasing the variation of spectral entropy values explored during normal sinus rhythm.
Last, windows straddling transitional periods of the heart rate
will also demonstrate atypical power spectra, further compounding the increase
in the variance when comparing normal sinus rhythm to atrial fibrillation.
% No variance difference for flutter
We do not conjecture on (and do not observe) a characteristic difference in the
variance of spectral entropy values for atrial flutter,
relying on the spectral entropy level to distinguish the arrhythmia
from fibrillation and normal sinus rhythm.

% Boundaries of disorder map
In theory, the spectral entropy can take values in the range [0,1].
Possible variances in sequences of spectral entropy values
then lie in the range [0,$\frac{1}{4}$].
These two ranges determine the two-dimensional cardiac disorder map.
In practice, we plot the standard deviation rather the variance for clarity,
and so rhythm thresholds are given in terms of the standard deviation.
Due to finite time and windowing considerations,
the spectral entropy is restricted to a subset of values within its possible range.
We attempt to find limits in the values that the spectral entropy can take
by applying the measure to synthetic event series: a periodic series with
constant interbeat interval, and a random series drawn from a Poisson
probability distribution with a mean firing rate.
For a heart rate range of 50 beats per minute (bpm) to 200 bpm in 1-bpm increments
we obtain 150 synthetic time series for the periodic and Poisson cases, respectively.
The average spectral entropy value over the 150 time series in the periodic case is 0.67$\pm$0.04;
the average value in the Poisson case is 0.90$\pm$0.01.
By assuming the maximum variance to occur in a rhythm that randomly changes between the
periodic and Poisson cases with equal probability, an approximate upper bound for the standard deviation
can be calculated:
using the two average spectral entropy values in the definition of the standard deviation,
we find the upper bound to be approximately 0.115.

% Getting disorder map
Figure \ref{Figure:Phase_Space}
illustrates the cardiac disorder map for all 25 patients comprising the afdb.
The standard deviation is calculated from $M$ adjacent spectral entropy values
(separated by \emph{a}), corresponding to a duration of $\beta=Ma=ML\tau/4$;
we call $\beta$ the variance window.
In this case, we have $M$ equal to 20 and so $\beta$ has a
length of 30 s for a typical patient.
We will see in the following section that $\beta$ sets the response time
of the arrhythmia detection algorithm.
The smallest useable number for $M$ is 4, corresponding to the rapid response
case where $\beta$ is typically 6 s.
We have $M$ equal to 40 for the case where $\beta$ is typically 60 s.
In Fig. \ref{Figure:Phase_Space}, each value of the standard deviation
is plotted against the average value of the spectral entropy within the variance window,
and is colored according to the rhythm assessment provided in the annotations.
As with spectral entropy windows, variance windows %, $\beta$,
have an overlap, \emph{b}.
For simplicity, we set \emph{b}=\emph{a}, giving a typical value of 1.5 s.
Note that \emph{b} can take any integer multiple of \emph{a},
though doing so does not alter the results substantially.

% Talk about phase space
One observes atrial fibrillation to be situated in the upper left of the disorder map,
consistent with having a high value for the spectral entropy and a low value
for the variance. Atrial flutter has a lower average value for
the spectral entropy, as expected.
% Get thresholds
For the given case with $\beta$ typically 30 s, we determine fibrillation
to exhibit spectral entropy levels above $\Gamma_{fib}$=0.84,
with flutter present below $\Gamma_{fl}$=0.70.
A standard deviation threshold can be inferred at around $\Phi_{fib}$=0.018,
with the majority of fibrillating points falling below that value.
Although beyond the expository purpose of this paper,
we note that these approximate thresholds can be further optimized using,
for example, discriminant analysis \cite{McLachlan}.
% Close out
Disorder maps for the three detection response times (6 s, 30 s, 60 s)
are qualitatively similar;
increasing the length of the variance window improves the separation of
rhythms in the disorder map at a cost of requiring more data per point.
From these observations, we hypothesize threshold values in the spectral entropy
level and variance that distinguish the two arrhythmias from normal sinus rhythm.
In the following section, thresholds drawn from the disorder map are used in an
arrhythmia detection algorithm.

\section{Algorithm}
\label{Algorithm}

In this section, we present a description of the automatic arrhythmia detection algorithm (Sec. \ref{ArrhythmiaDetectionAlgorithm}),
followed by results for a range of detection response times (Sec. \ref{Results}).

\subsection{Arrhythmia detection algorithm}
\label{ArrhythmiaDetectionAlgorithm}

The arrhythmia detection algorithm uses
thresholds in the level and variance of spectral entropy values observed
in the cardiac disorder map to automatically detect and label rhythms
in patient event series data.
% Why no flutter
The afdb contains significantly fewer periods of atrial flutter compared to
atrial fibrillation and normal sinus rhythm
(periods of flutter total 1.27 h, whereas periods of fibrillation total 91.59 h),
the typical length of periods of flutter is of the order tens of seconds.
Of the eight patients annotated as having flutter, only patients 04936 and 08378
have periods of flutter long enough (i.e., $>\beta$) for analysis by the algorithm.
For this reason we do not
include here the flutter prediction method of the algorithm,
although extensions including flutter follow a similar principle and
are simple in practice to implement.
Other studies using the afdb (e.g., \cite{TatenoGlass})
restrict themselves to methods differentiating only between fibrillation and
normal sinus rhythm.
Additional comments on the practicality of detecting atrial flutter and
selected results for flutter
will be given in the Discussion section (Sec. \ref{DisagreementswithAnnotations}).

The five stages of the algorithm are shown in Fig. \ref{Figure:Schematic}.
The first three stages have been covered in depth as part of the Data Analysis
section, but we include a brief summary here for completeness.
We first obtain a binary string representing the dynamics of the heart
for a given patient by discretizing the physionet data every $\tau$=30 ms
(stage 1 to stage 2).
In stage 3, the spectral entropy measure is applied for windows
of duration $\alpha=L\tau$, with $L$ chosen for each patient such that
there are on average ten beats within the spectral entropy window,
giving $\alpha$ as 6 s for a typical patient.
Using an overlap parameter \emph{a} (typically 1.5 s), leads to a series of
spectral entropy values separated in time by this amount.
% Prediction
Given no prior knowledge of the provided rhythm assessments, we calculate
the standard deviation and average magnitude of $M$ spectral entropy values in
variance windows of length $\beta=Ma$ preceding a given time point.
We use the example case of $M$ equal to 20 (giving $\beta$ as 30 s for a typical patient).
The level and standard deviation thresholds for atrial fibrillation are set consistent
with values obtained from the cardiac disorder map, for this case we determine
$\Gamma_{fib}$=0.84 and $\Phi_{fib}$=0.018.
Stage 4 generates preliminary predictions for the rhythm state of the heart:
we denote as fibrillating (AF) instances where the spectral entropy level
is greater than $\Gamma_{fib}$ and the standard deviation is less than $\Phi_{fib}$,
with all other combinations considered to be normal sinus rhythm (N) \cite{normal_comment}.
Setting the overlap of variance windows such that \emph{b} = \emph{a},
we obtain a string of rhythm predictions drawn from the set \{AF, N\}
and separated in time by \emph{b}.

% Modal smoothing
Finally, in stage 5 we apply a rudimentary smoothing procedure
to the initial string of rhythm predictions.
For a particular prediction, we consider a preceding period
$\gamma=2\beta+\emph{b}=(2M+1)L\tau/4$,
leading in this example to a typical length for $\gamma$ of 61.5 s.
We find the \emph{modal} prediction: the prediction \{AF, N\}
occurring most frequently in $\gamma$,
labeling the modal prediction \{AF$^{\prime}$, N$^{\prime}$\}.
We call $\gamma$ the modal smoothing window.
In this form, we understand the windows $\beta$ and $\gamma$ as setting the
response time of the algorithm: $\beta$ is defined in terms of the number of
preceding spectral entropy values required for a given prediction;
for $\gamma$ to register a change in rhythm, over half of the predictions
must suggest the new rhythm.
The response time is then $\frac{\gamma}{2}$, which is approximately equal
to $\beta$.
We have the modal smoothing windows overlapping with parameter
\emph{c}=\emph{b}=\emph{a}.
This results in a final time series of predictions and constitutes the
output of the arrhythmia detection algorithm for a given patient.
An example of the algorithm output for patient 08378
(including a threshold for atrial flutter)
is shown in Fig. \ref{Figure:Patient_08378}.

% Outro
We apply the above steps, comprising the three data windows ($\alpha$, $\beta$, $\gamma$),
to each patient in the afdb.
Specifying $\tau$, $L$ and $M$ fixes the remaining parameters, their exact magnitude
determined by $L$.
A summary of windowing symbols can be found in Table \ref{taba}.
Values for the atrial fibrillation threshold parameters ($\Gamma_{fib}$ and $\Phi_{fib}$)
are kept the same for each patient for a given response time.
The results obtained from the algorithm are described in the following
section.

\subsection{Algorithm results}
\label{Results}

% Results text
We now present the results of the cardiac arrhythmia detection algorithm
for atrial fibrillation.
The following window parameters were used:
$\tau$ is set to 30 ms, $L$ is chosen such that $\alpha$ is expected to contain
10 beats, and $M$ is set to 20, windows have overlap parameters
\emph{c}=\emph{b}=\emph{a}=$\alpha /4$
(for typical patients in the afdb,
$\alpha\sim6s$, $\beta\sim30s$, $\gamma\sim61.5s$, and $a\sim1.5s$).
Threshold values for fibrillation are set at
$\Gamma_{fib}$=0.84 for the spectral entropy level and $\Phi_{fib}$=0.018
for the standard deviation.
Each prediction produced by the algorithm (denoted by a primed symbol)
is compared with the rhythm assessment documented in the database and can
be classified into one of four categories \cite{Hulley}:
true positive (TP), AF is classified as AF$^{\prime}$;
true negative (TN), non-AF is classified as non-AF$^{\prime}$;
false negative (FN), AF is classified as non-AF$^{\prime}$;
false positive (FP), non-AF is classified as AF$^{\prime}$.
Percentages of these quantities for each patient and for the entire afdb
are given in Table \ref{tab1}.
Overall, we obtain a predictive capability
(assessed using the percentage of predictions agreeing with the provided annotations) of 89.5\%.
The sensitivity and specificity metrics are defined by TP/(TP+FN) and TN/(TN+FP), respectively.
The predictive value of a positive test (PV$_+$) and the predictive value of
a negative test (PV$_-$) are defined by TP/(TP+FP) and TN/(TN+FN), respectively.
These, and results for other values of $\beta$ are given in Table \ref{tab2}.

% Other results
In repeating the algorithm with different values for the variance window,
shorter $\beta$ represents a quicker response time.
We obtain for each $\beta$ a new disorder map to determine the
relevant threshold values.
For the rapid response case, $\beta$ typically 6 s, we alter the
fibrillating thresholds in the arrhythmia detection algorithm to be
$\Gamma_{fib}$=0.855 and $\Phi_{fib}$=0.016;
we find a predictive capability of 85.7\%.
With $\beta$ typically 60 s, the fibrillating thresholds become
$\Gamma_{fib}$=0.84 and $\Phi_{fib}$=0.019;
the predictive capability is 90.3\%.

% TABLE: RESULTS
\begin{table}[t]
\centerline{
\begin{tabular}{|c|cc|cc|}
\hline
Patient&TP (\%)&TN (\%)&FN (\%)&FP (\%)\\
\hline
00735&0.8&85.0&0.0&14.2\\
03665&29.8&30.4&37.8&2.0\\
04015&0.5&92.4&0.2&6.9\\
04043&8.9&76.5&13.1&1.5\\
04048&0.4&98.8&0.7&0.1\\
04126&3.3&78.3&0.6&17.8\\
04746&53.6&43.8&0.8&1.8\\
04908&7.0&88.2&1.6&3.2\\
04936&43.1&19.0&36.3&1.6\\
05091&0.0&85.6&0.2&14.2\\
05121&56.9&30.5&8.4&4.2\\
05621&0.9&94.9&0.4&3.8\\
06426&92.7&1.9&3.1&2.3\\
06453&0.4&97.7&0.7&1.2\\
06995&42.8&47.1&3.0&7.1\\
07162&100.0&0.0&0.0&0.0\\
07859&83.1&0.0&16.9&0.0\\
07879&53.3&38.1&8.5&0.1\\
07910&13.5&85.7&0.5&0.3\\
08215&80.0&19.7&0.3&0.0\\
08219&18.3&59.8&3.8&18.1\\
08378&20.0&77.3&2.4&0.3\\
08405&68.9&28.4&2.7&0.0\\
08434&3.8&91.6&0.2&4.4\\
08455&65.6&31.5&2.9&0.0\\
\hline
Total&36.1&53.4&6.5&4.0\\
\hline
&True:&89.5\%&False:&10.5\%\\
\hline
\end{tabular}}
\caption{Results of the arrhythmia detection algorithm
using data in the MIT-BIH atrial fibrillation database.
For the parameters used, see Algorithm results section (Sec. \ref{Results}).
Algorithm predictions (primed symbols) are compared to annotated rhythm
assessments.
TP, AF is classified as AF$^{\prime}$;
TN, non-AF is classified as non-AF$^{\prime}$;
FN, AF is classified as non-AF$^{\prime}$;
FP, non-AF is classified as AF$^{\prime}$.
}
\label{tab1}
\end{table}

% TABLE: RESULTS SUMMARY
\begin{table}[t]
\centerline{
\begin{tabular}{|cc|c|cccc|}
\hline
$M$&$\beta$&True (\%)&Sens. (\%)&Spec. (\%)&PV$_+$ (\%)&PV$_-$ (\%)\\
\hline
4&6s&85.7&82.1&88.4&83.9&87.0\\
20&30s&89.5&84.8&92.9&89.8&89.2\\
40&60s&90.3&83.6&95.2&92.8&88.7\\
\hline
\end{tabular}}
\caption{Summary of results for variance windows of different lengths.
Length is set by parameter $M$=4, 20, 40, giving durations for typical patients: $\beta\sim$ 6s, 30s, 60s, respectively.
Shorter $\beta$ indicates a quicker response time.
Metrics defined as,
sensitivity,  TP/(TP+FN);
specificity, TN/(TN+FP);
PV$_+$, TP/(TP+FP);
PV$_-$, TN/(TN+FN).
}
\label{tab2}
\end{table}

\section{Discussion}
\label{Discussion}

% Section overview
We begin with an exposition of the results presented in the previous section
and the effects of different parameter values on the output of the arrhythmia
detection algorithm.
This is followed by a discussion, with reference to the electrocardiograms
provided as part of the afdb, of disagreements between the provided
rhythm annotations, measures relying solely on the heart rate, and the
predictions of our algorithm (Sec. \ref{DisagreementswithAnnotations}).
Having shown that some of the annotations may be unreliable, we comment on
situations where the algorithm may still present incorrect predictions
(Sec. \ref{OtherRhythms}).
The benefits of the spectral entropy measure compared to other
fibrillation detection methods is then given (Sec. \ref{ComparisontoOtherMethods}).
We close the section with a discussion of the systematic windowing errors present
in our procedure (Sec. \ref{SystematicError}).

% Summarize results
Instances of atrial fibrillation constitute approximately 40\% of the afdb.
If we consider a null-model where we constantly predict normal sinus rhythm,
we would expect a predictive capability of around 60\%.
In Table \ref{tab2}, we observe an improvement in the predictive capability
of the detection algorithm when the length of the variance window, $\beta$,
is increased from 6 s (85.7\%) to 60 s (90.3\%) for a typical patient.
The choice of shorter $\beta$ improves the response time of the algorithm by
requiring less data per prediction; values for $\beta$ less than 6 s do not
incorporate enough data to give meaningful results.
Increasing $\beta$ beyond 30 s improves the predictive capability very little.
This suggests that additional factors, independent of the specific
parameters chosen here, need to be considered.
Results in Table \ref{tab1} for the case
$\beta$ typically 30 s indicates an overall predictive capability
of 89.5\%. For individual patients, the predictive capability
ranges from 60.2\% (patient 03665) to 100\% (patient 07162).
To explain this variation, we investigate the
form of patient ECGs during periods of disagreement between annotation
and prediction.
% REF TO SUPPLEMENTARY INFORMATION
Examples of the ECGs referred to in
Secs. \ref{DisagreementswithAnnotations} and \ref{OtherRhythms}
are included in the supplementary information that accompanies
this paper \cite{Supplementary_Info}.

\subsection{Disagreements with annotations}
\label{DisagreementswithAnnotations}

% Disagreements with annotations
Rhythm assessments have been questioned before \cite{TatenoGlass};
here, we give explicit examples where we believe the
ECGs to suggest a rhythm different from that given by the annotation.
We observe in the ECGs of patients 08219 and 08434 periods of
atrial fibrillation that we believe to have been missed in the annotations
but are correctly identified by our detection algorithm \cite{08219_08434}.
Cases such as these serve to negatively impact the results of the algorithm unfairly;
however, we note that such instances comprise a small proportion of the afdb.
% Comments on flutter
Atrial flutter may have been misannotated in patients 04936 and 08219 \cite{04936_08219};
in particular, two considerable periods of flutter may have been annotated incorrectly in patient 04936.
This unreliability of rhythm assessment, compounded with the limited number
of periods of atrial flutter in the database, prevents us from drawing meaningful
quantitative conclusions regarding the success of the detection algorithm in identifying flutter.
Despite this, we believe that the spectral entropy is in principle still capable of
identifying flutter (see Fig. \ref{Figure:Patient_08378}).
Returning to the two patients with significant periods of flutter,
we run the algorithm with the inclusion of a threshold for atrial flutter
motivated by each patient's individual disorder map: $\Gamma_{fl}$
(other parameters as per the Algorithm results section with $M=20$).
For patient 08378 with $\Gamma_{fl}=0.70$, we find 86.3\% agreement with the annotations for flutter;
for patient 04936 with $\Gamma_{fl}=0.81$, we find 66.9\% agreement, bearing in mind the points raised above.

% Comments on HR
Consideration of ECGs demonstrates the inability of measures
relying solely on the heart rate and its derivatives to consistently
distinguish between fibrillation, flutter and other rhythms.
Atrial fibrillation is characteristically associated with an
elevated heart rate (100-200 bpm) \cite{Bennett};
atrial flutter exhibits an even higher heart rate ($>$150 bpm) with a
sharp transition from normal sinus rhythm.
This expected behavior, whilst found to hold qualitatively for the majority
of patients, fails during large periods for patient 06453 and is completely
reversed for patient 08215 \cite{06453_08215}.
The resting heart rate is also found to differ dramatically between patients in the afdb.
The spectral entropy, being less susceptible to variations in the heart rate \cite{Cammarota},
is better suited to form the basis of a detection algorithm compared to a
measure relying solely on heart rate.

\subsection{Other rhythms}
\label{OtherRhythms}

% Where algorithm can still get confused (P61)
% False negatives
The unreliability of parts of the annotations still does not account for all
false predictions produced by the detection algorithm.
We suggest the presence of other rhythms within the afdb to be an additional factor
that needs to be considered.
Table \ref{tab2} shows the sensitivity metric to be consistently lower
for all values of $\beta$, suggesting a bias towards false negatives
(FNs occur when AF is classified as non-AF$^{\prime}$).
FNs total 6.5\% for $\beta$ typically 30 s in Table \ref{tab1},
and comprise 36.3\% of predictions for patient 04936.
Given our requirement in the detection algorithm for periods that are classed
as AF to satisfy both a spectral entropy level \emph{and} variance condition,
FNs are most likely to arise when one threshold condition fails to be met.
Cases where the variance threshold is not satisfied may be associated with
the physiological phenomena of \emph{fib-flutter} and \emph{paroxysmal atrial fibrillation},
and would be located right of the standard deviation threshold on the
disorder map (Fig. \ref{Figure:Phase_Space}).
Fib-flutter corresponds to periods where the rhythm transitions in quick succession
between atrial fibrillation and flutter \cite{Kadish}, with paroxysmal fibrillation describing
periods where atrial fibrillation stops and starts with high frequency.
Such behavior naturally causes the variance to increase and one might question
whether it is still appropriate to classify those periods as standard atrial fibrillation.
We identify in the ECG of patient 04936 periods of fib-flutter which likely
accounts for the high proportion of FN results \cite{04936};
by inspecting the patient's disorder map, we indeed observe points annotated
as atrial fibrillation with uncharacteristically high standard deviation, signifying that
fib-flutter would be a more accurate rhythm classification.
Cases where the spectral entropy level threshold is not met can occur
when QRS complexes indicative of atrial fibrillation appear with unusually
regular rhythm; such behavior would lie below the level threshold on the disorder map.
Owing to the small number of beats contained within each window, such
occurrences inevitably arise; the process of modal smoothing lessens the impact
of this phenomenon in the arrhythmia detection algorithm.

% False positives
False positives (FPs occur when non-AF is classified as AF$^{\prime}$),
which comprise 4.0\% of the afdb for $\beta$ typically 30 s, may also have a
physiological explanation.
During \emph{sinus arrhythmia}, there are alternating periods of slowing and
increasing node firing rate, while still retaining QRS complexes indicative of
normal sinus rhythm. These alternating periods increase the irregularity of
beats within the spectral entropy window. If the variance threshold is also satisfied,
sinus arrhythmia may be incorrectly classified as AF$^{\prime}$ by the
arrhythmia detection algorithm.
\emph{Sinus arrest} occurs when the sinoatrial node fails to fire and results in
behavior that is similar in principle to sinus arrhythmia; these two conditions are
likely responsible for the high proportion of FPs (14.2\%) that are observed
in patient 05091 \cite{05091}.

\subsection{Comparison to other methods}
\label{ComparisontoOtherMethods}

% We are not trying to predict atrial fibrillation
Vikman \emph{et al.} showed that decreased ApEn values of
heart beat fluctuations have been found to precede
(at timescales of the order an hour) spontaneous episodes
of atrial fibrillation in patients without structural
heart disease \cite{Vikman}.
We stress that the algorithm presented here is not
intended to predict in advance occurrences of fibrillation;
rather, it is designed to detect the onset of fibrillation
as quickly as possible using only interbeat intervals.
% Benefits / comparison to T&G
Tateno and Glass \cite{TatenoGlass} present an atrial fibrillation detection
method that is statistical in principle and based upon an observed difference in the
standard density histograms of \emph{$\Delta$RR} intervals
(the difference in successive interbeat intervals).
A series of reference standard density histograms characteristic of
atrial fibrillation (as assessed in the annotations) are first obtained from the afdb.
Their detection algorithm is re-run on the afdb by taking 100 interbeat intervals
and comparing them to the reference histograms, where appropriate predictions can then be made.
The reference histograms rely on the correctness of the annotations
in order to determine fibrillation, whereas the thresholds in our algorithm
are only weakly dependent on the data set under consideration.
% Sample-training separately bit
Figure \ref{Figure:Phase_Space} is an empirical observation,
in future analyses we would like to use fibrillation thresholds derived from a data set separate
from the one under consideration.

Sarkar \emph{et al.} have developed a detector of atrial fibrillation and tachycardia that
uses a Lorentz plot of \emph{$\Delta$RR} intervals to differentiate between rhythms \cite{Sarkar}.
The detector is shown to perform better for episodes of fibrillation greater
than 3 min and has a minimum response time of 2 min.
By contrast, our method is applicable to short sections of data, enabling quicker
response times to be used.
We see our algorithm complementing other detection techniques,
with the potential for an implementation that combines more than one method.
Combining methods becomes increasingly relevant when running algorithms on
data sets containing a variety of arrhythmias.
As noted in \cite{TatenoGlass}, other arrhythmias often show irregular \emph{RR} intervals,
and previous studies have found difficulty in detecting atrial fibrillation based solely
on \emph{RR} intervals \cite{Andresenplusothers}.

\subsection{Systematic error}
\label{SystematicError}

% Error
There are two intrinsic sources of error in the spectral entropy measure
related to the phenomenon of \emph{spectral leakage}:
that due to the ``picket-fence effect" \cite{Salvatore}
(where frequencies in the power spectrum fall between discrete bins)
and that due to finite window effects \cite{Harris}
(where, for a given frequency, an integer number of periods
does not fall into the spectral entropy window).
% Quantify error
We attempt to quantify this error by applying the measure (with
parameters as per the Data Analysis section) to synthetic event series:
a periodic series with constant interbeat interval.
For a heart rate range of 50-200 bpm in 1-bpm increments
we obtain 150 synthetic time series.
We find the average error in the spectral entropy
over the 150 time series to be 0.02.
% Error in variance
The average standard deviation value
(with variance windows having $M$ equal to 20 spectral entropy values)
over the 150 time series is 0.011$\pm$0.009;
the average error on these standard deviation values due to
windowing is 0.0002.

% Discussion / minimization of errors
The presence of some form of error associated with finite windows is
unavoidable. We have attempted to minimize such errors by choosing
parameters that achieve a balance between usability and error magnitude.
There is still scope for fine-tuning parameters - in particular, trying a
variety of window shapes to further reduce the affect of spectral leakage.
However, we find the general results to be robust to a range of window parameters,
implying any practical effect of windowing errors to be minimal when compared to
the other issues discussed in this section.

\section{Further Work}
\label{FurtherWork}

Additional directions for this work include refining and extending
our cardiac study with a view to clinical implementation.
Furthermore, we suggest that rhythmic signals arising from other
biological systems may have application for the techniques described
in this paper.
% Finding optimum parameter set
An investigation of the optimal windowing parameter set would be
instructive since our findings suggest the existence of physiological
thresholds in the spectral entropy level and variance that are
applicable to a variety of patients.
% MIT-BIH arrhythmia database (P65)
As noted at the end of Sec. \ref{ComparisontoOtherMethods}, one challenge
would be to investigate and improve the utility of the measure (alone or combining methods)
when applied to patients that demonstrate a mix of different pathologies and arrhythmias.
Adjusting the spectral entropy window to covary with instantaneous heart rate
so that $\alpha$ always contains ten beats exactly would further reduce issues
related to variations in the heart rate.
Extending the algorithm to include other
dimensions in the disorder map (e.g., heart rate) will likely improve
the accuracy of results and may increase the number of arrhythmias the
spectral entropy can distinguish between.

% Clinical
An accurate automatic detector of atrial fibrillation
would be clinically useful
in monitoring for relapse of fibrillation in patients and in assessing
the efficacy of antiarrhythmic drugs \cite{Hohnloser}.
An implementation integrated with an ambulatory ECG or heart rate monitor
would be useful in improving the understanding of arrhythmias on
time scales longer than that available using conventional
ECG analysis techniques alone.

% Application to other biological systems
Measures of disorder in the frequency domain have practical significance in
a range of biological signals. For example, the regularity of the background
electroencephalography (EEG is the measurement of electrical activity produced
by the brain as recorded from electrodes placed on the scalp)
alters with developmental and psychophysiological factors:
some mental or motor tasks cause localized desynchronization; in addition,
the background becomes more irregular in some neurological and psychiatric
disorders (\cite{Inouye} and references therein).
The spectral entropy method and the concept of the disorder map
described in this paper are not cardiac specific:
it would be instructive to adapt these ideas to other rhythmic signals
where a rapid detection of arrhythmia would be informative.

\section{Conclusion}
\label{Conclusion}

% Summary
In this paper we have presented an automatic arrhythmia detection algorithm
that is able to rapidly detect the presence of atrial fibrillation
using only the time series of patients' beats.
The algorithm employs a general technique for quickly quantifying
disorder in high-frequency event series:
the spectral entropy is a measure of disorder applied to the power spectrum
of periods of time series data.
The physiologically motivated use of the spectral entropy
is shown to distinguish atrial fibrillation and flutter from other rhythms.
For a given set of parameters, we are able to determine from a disorder map
two threshold conditions (based on the level and variance of spectral entropy values)
that enable the detection of fibrillation in a variety of patients.
We apply the algorithm to the MIT-BIH atrial fibrillation database of 25 patients.
When the algorithm is set to identify abnormal rhythms within 6 s
it agrees with 85.7\% of the annotations of professional rhythm assessors;
for a response time of 30 s this becomes 89.5\%, and with 60 s it is 90.3\%.
The algorithm provides a rapid way to detect fibrillation,
demonstrating usable response times as low as 6 s
and may complement other detection techniques.
There also exists the potential for our spectral entropy and disorder map
implementations to be adapted for the rapid identification of disorder in
other rhythmic signals.

\begin{acknowledgements}
P.P.A.S. thanks the CABDyN Complexity Centre (Oxford)
and the Centre of Excellence in Computational Complex Systems Research (Helsinki);
C.F.L. thanks the Glasstone Trust (Oxford) and Jesus College (Oxford);
N.S.J. thanks the EPSRC and BBSRC for support, and the Rey Institute for
Nonlinear Dynamics in Medicine.

\end{acknowledgements}

% BIBLIOGRAPHY

% to cite, use command \cite{...reference tag...}
% to obtain bold font, use the command: {\bf number}


\begin{thebibliography}{30}
\bibitem{normal_comment} More correctly, we mean nonfibrillating,
but we find the overwhelming majority rhythm to be normal sinus rhythm.
\end{thebibliography}
\end{document}